\documentclass[graybox]{svmult}
\usepackage{cite}
\usepackage{soul}
\usepackage{braket}
\usepackage{type1cm}        
\usepackage{makeidx}         
\usepackage{graphicx}        
\usepackage{multicol}        
\usepackage[bottom]{footmisc}

\usepackage{newtxtext}       %
\usepackage{newtxmath}       

\makeindex             

\usepackage[qm]{qcircuit}
\usepackage{textcomp}
\usepackage{subcaption, wasysym}
\usepackage{url}

\begin{document}

\title*{Architectures for Quantum Information Processing}


\author{Suryansh Upadhyay, Mahabubul Alam, and Swaroop Ghosh}
\institute{
    Suryansh Upadhyay \at Pennsylvania State University, University Park, PA 16802, USA \email{sju5079@psu.edu}
    \and Mahabubul Alam \at Pennsylvania State University, University Park, PA 16802, USA \email{mxa890@psu.edu}
    \and Swaroop Ghosh \at Pennsylvania State University, University Park, PA 16802, USA \email{szg212@psu.edu}
}
%
%
\maketitle
\abstract{
Quantum computing is changing the way we think about computing. Significant strides in research and development for managing and harnessing the power of quantum systems has been made in recent years, demonstrating the potential for transformative quantum technology. Quantum phenomena like superposition, entanglement, and interference can be exploited to solve issues that are difficult for traditional computers. IBM's first public access to true quantum computers through the cloud, as well as Google's demonstration of quantum supremacy, are among the accomplishments. Besides, a slew of other commercial, government, and academic projects are in the works to create next-generation hardware, a software stack to support the hardware ecosystem, and viable quantum algorithms. This chapter covers various quantum computing architectures including many hardware technologies that are being investigated. We also discuss a variety of challenges, including numerous errors/noise that plague the quantum computers. An overview of literature investigating noise-resilience approaches is also presented. }

\begin{figure}
    \centering
    \includegraphics[width=5in]{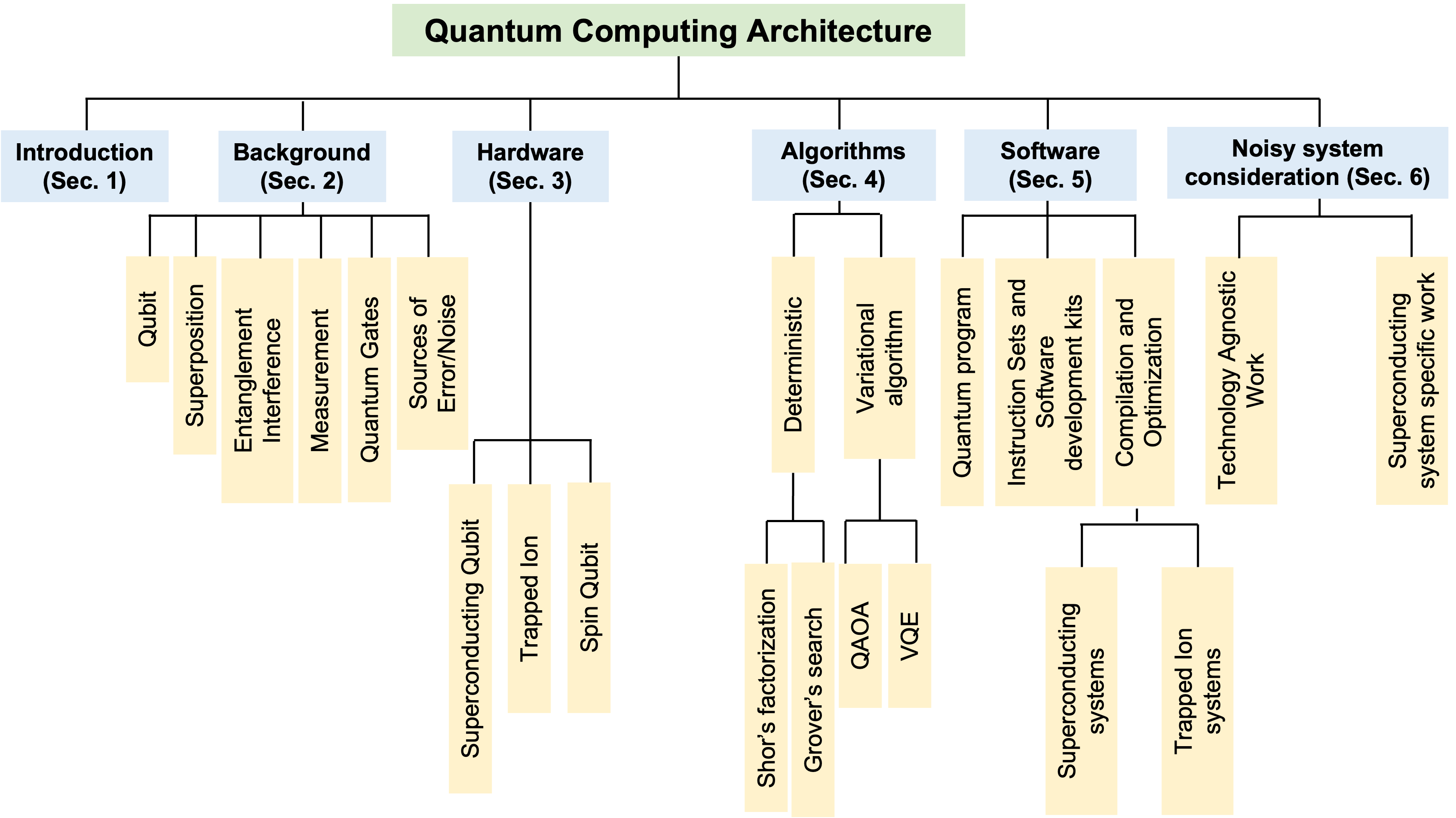}
    \caption{Chapter structure}
    \label{7}
\vspace{-4mm}
\end{figure}

\section{Introduction}
\label{sec:intro}

Quantum Computing (QC) exploits phenomena such as superposition, entanglement, and interference to efficiently explore exponentially large state spaces and compute solutions for certain classically intractable problems. Despite the extraordinary development in the classical computing domain, including device, architecture, and algorithm in past decades, many problems cannot be solved in a reasonable time, even on the fastest classical supercomputer. For example, solving the nitrogen fixation problem on the best classical supercomputer would take more than 2 million years \cite{1}. Since 1982, when Richard Feynman originally envisioned the quantum-mechanical computer, researchers have taken incremental but significant efforts toward developing quantum algorithms and hardware. The original Church-Turing thesis \cite{86} in it's contrapositive form states that a computation that cannot be performed by a Turing machine cannot be performed without breaking a physical law. It is a physics principle because it implies a limitation on what physical systems can do. However, quantum machines are able to perform calculations in polynomial time that Turing machines are believed to not be able to do, e.g., factorization. David Deutsch and Richard Jozsa presented the ``Deutsch-Jozsa algorithm" \cite{2}, which illustrated how a quantum algorithm might perform a job in fewer steps than a conventional version. Shor's algorithm\cite{3}, which is one of the most well-known quantum algorithms, can factor an integer in prime numbers tenfold quicker than the best conventional solution. The impact of this exponential speedup on encryption and internet security is substantial. Cirac and Zoller suggested the experimental implementation of the Controlled NOT (CNOT) gate \cite{4} on the hardware side. Following that, nuclear magnetic resonance (NMR)-based devices were used to demonstrate quantum computing hardware, including the first demonstration of a full-fledged quantum algorithm on a 2-qubit NMR computer at Oxford University. Schoelkopf, Devoret, Girvin, and colleagues at Yale University invented the superconducting ``Transmon" qubit, which revolutionized qubit technology and paved the road for scalability. IBM first provided access to a 5-qubit programmable quantum computer through the cloud. The free access to quantum computers attracted many researchers around the globe to the world of quantum computing. The quantum threshold theorem \cite{87} demonstrated that if the error to perform each gate is a small enough, one can perform arbitrarily long quantum computations to arbitrarily good precision with only a small increase in gate count. This shows that quantum computers can be made fault-tolerant. Especially in the last few years, quantum computing has experienced breakthroughs across the stack, including algorithm, architecture, and hardware. The most notable of these is the demonstration of Quantum Supremacy by Google \cite{5}. The group of researchers from Google performed a task on a 53-qubit quantum computer in seconds which arguably would take days on the fastest classical supercomputer. Application domains of QC now include machine learning \cite{6}, security \cite{7}, drug discovery \cite{8}, computational quantum chemistry \cite{9} and optimization \cite{10}. 

On one hand, researchers are proposing new quantum algorithms to speed up computation, while on the other hand, various technologies like superconducting, trapped-ion (TI), and photonics are also being studied to design efficient quantum bits or qubits.
Despite all the signs of progress, quantum computers are yet to solve a practical-scale problems.

The quantum computer architecture is essential for the functioning of a quantum computer. Despite various system optimizations, the performance of a quantum processor is still severely constrained by the amount of available computation resource. This chapter covers the fundamentals of quantum computing architectures including some of the most often used hardware architectures and the associated software stacks. We discuss the numerous issues that these technologies face, as well as a literature review of efforts to address these architectural issues for both hardware and software stacks. Fig. \ref{7} illustrates the topics covered in this chapter.

\section{Background}
\label{sec:basics}
This section gives a quick introduction of quantum computing's fundamental ideas in order to provide readers an understanding of the new computing paradigm allowed by quantum properties and the associated challenges. 
\subsection{Quantum Bits (Qubits)}
\begin{figure}
    \centering
    \includegraphics[width=4in]{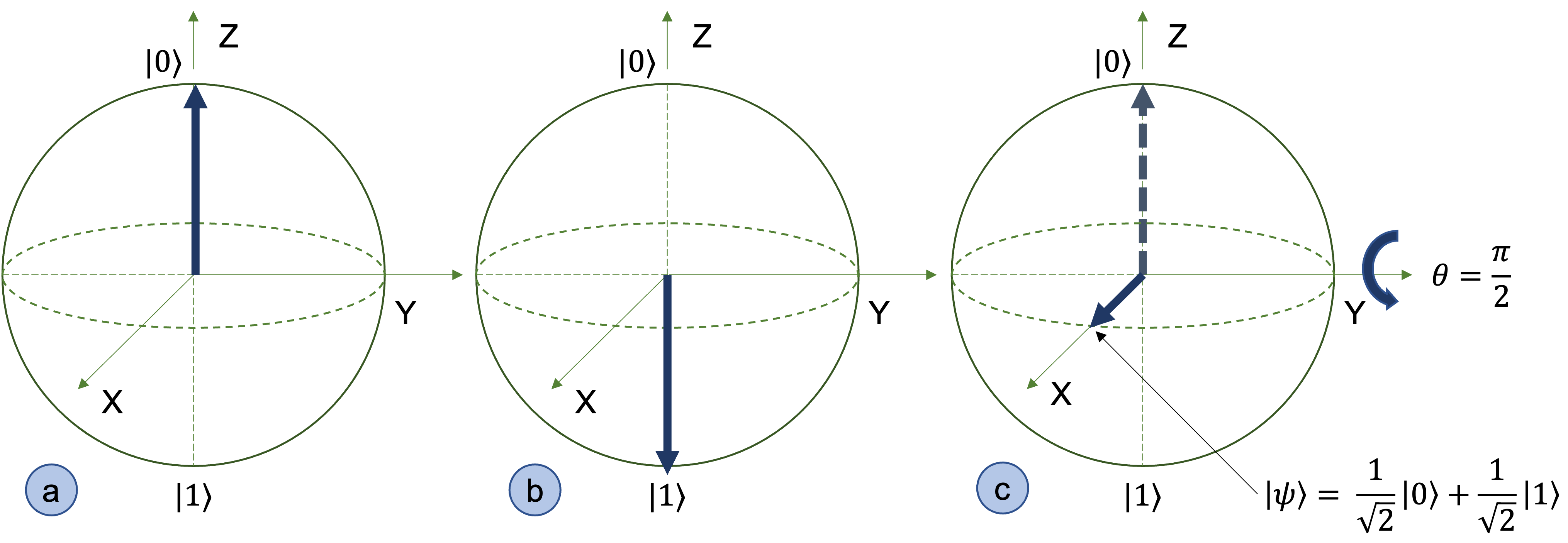}
    \caption{Bloch sphere representation of state a) $\ket{0}$ and state b) $\ket{1}$.c) Bloch sphere representation of the R$_{Y}(\pi/2)$ gate on state $\ket{0}$.  }
    \label{1}
\vspace{-4mm}
\end{figure}

In quantum computing, a qubit is the basic unit of quantum information, the quantum version of the classic binary bit physically realized with a two-state device. For example, electron spin can realize a qubit where electron up-spin can represent data ‘1’ and down-spin can represent data ‘0’. Therefore, a qubit has two quantum states, analogous to the classical binary states. However, classical bit can be either ‘0’ or ‘1’ at a time, qubits can be a mixture of both 0 and 1 simultaneously due to quantum superposition. Mathematically, a qubit state is represented by state vector $\psi = a \ket{0} + b \ket{1}$ where $|a|^2$ and $|a|^2$ represent probabilities of ‘0’ and ‘1’ respectively. Suppose a qubit is in state $\psi = .707 \ket{0} + .707 \ket{1}$. Reading out this multiple time will theoretically generate 0s and 1s with equal probability (as 0.7072 = 0.5). The state of a qubit is often represented visually by a Bloch sphere (Fig. \ref{1}), where the poles of the sphere represents states 0 and 1, and the equator is the perfect superposition.
Furthermore, qubit states can be entangled, allowing two or more qubit states to be correlated. The states of other entangled qubits can be changed by performing a single operation on one of the entangled qubits. Quantum interference is a property of qubit states that algorithm designers can take advantage of. A qubit state's amplitude can be both positive and negative. As a result, a quantum algorithm designer can tweak the gate operations so that a negative amplitude of the same qubit state cancels out a positive amplitude. Quantum superposition, entanglement, and interference, according to researchers, are at the heart of quantum speed-ups of quantum algorithms. Upon measurement, the qubit’s coefficient (or amplitude) becomes one in the state that is read and zero in the other; all information about the amplitudes is destroyed upon measurement also known as collapse of state.

\subsection{Quantum gates}
\label{subsec:qgates}
Gates are used in quantum computing systems to regulate qubit amplitudes and execute computations. At any given time, gates can act on one or more qubits. QC systems often support a set of universal single-qubit and two-qubit gates, similar to universal gates in classical computing. Quantum gates, unlike classical logic gates, are not physically formed; instead, they are realized through the use of pulses. These gate sets are used to express QC applications. A sequence of gates is executed on a set of correctly initialized qubits to run a program. The gates change the amplitudes of the qubits, moving the state space closer to the desired output. Intuitively, the gate pulses cause distinct rotations along different axes in the Bloch sphere (depending on pulse amplitude, duration, and shape). For example, the R$_{Y}(\pi/2)$ (rotation along Y-axis) can be a quantum gate, and it will rotate a qubit state by $\pi/2$ radian around Y-axis (e.g., applying an R$_{Y}(\pi/2)$ will put a qubit in $\ket{0}$ state in the superposition state Fig. \ref{1}c).

Mathematically, quantum gates are represented using unitary matrices (a matrix U is unitary if UU$^\dagger$ = I, where U$^\dagger$ is the adjoint of matrix U and I is the identity matrix). For an n-qubit gate, the dimension of the unitary matrix is 2n×2n. Any unitary matrix can be a quantum gate. However, in existing systems, only a handful of gates are possible, often known as the native gates or basis gates of that quantum processor. For IBM systems, the basis gates are ID, RZ, SX, X, and CNOT. CNOT is the only 2-qubit gate, and others are single-qubit. Any non-native gate in a quantum circuit is first decomposed using the native gates.

\subsection{Quantum error}
\label{subsec:qerror}

Quantum systems are plagued with noise because of the quantum gates being error prone. Besides, the qubits suffer from decoherence i.e., the qubits spontaneously interact with the environment and lose states. Therefore, the output of a quantum circuit is erroneous. The deeper quantum circuit needs more time to execute and gets affected by decoherence. More gates in the circuit also increase the accumulation of gate error. Parallel gate operations on different qubits can affect each other’s performance which is known as crosstalk. In this section we elaborate on these errors:
\subsubsection{Gate error}
Quantum gates are realized with pulses, and the pulses can be erroneous. For example, consider the R$_{Y}(\pi/2)$ gate. Due to variation, the pulse intended for a $\pi/2$ rotation may not result in an exact $\pi/2$ rotation, and it may under-rotate or over-rotate, leading to erroneous logical operation. As a result, gate failures are caused by faulty logical operations. For present systems, 2-qubit gate errors (e.g., CNOT error) are an order of magnitude larger than 1-qubit gate faults. A quantum circuit with a larger number of gates will accrue more gate faults, lowering the quantum program's reliability. Hence, compilation and error-tolerant mapping have the goal of reducing the number of gates in the quantum circuit.
\subsubsection{Relaxation and Dephasing}
\begin{figure}
    \centering
    \includegraphics[width=3.5in]{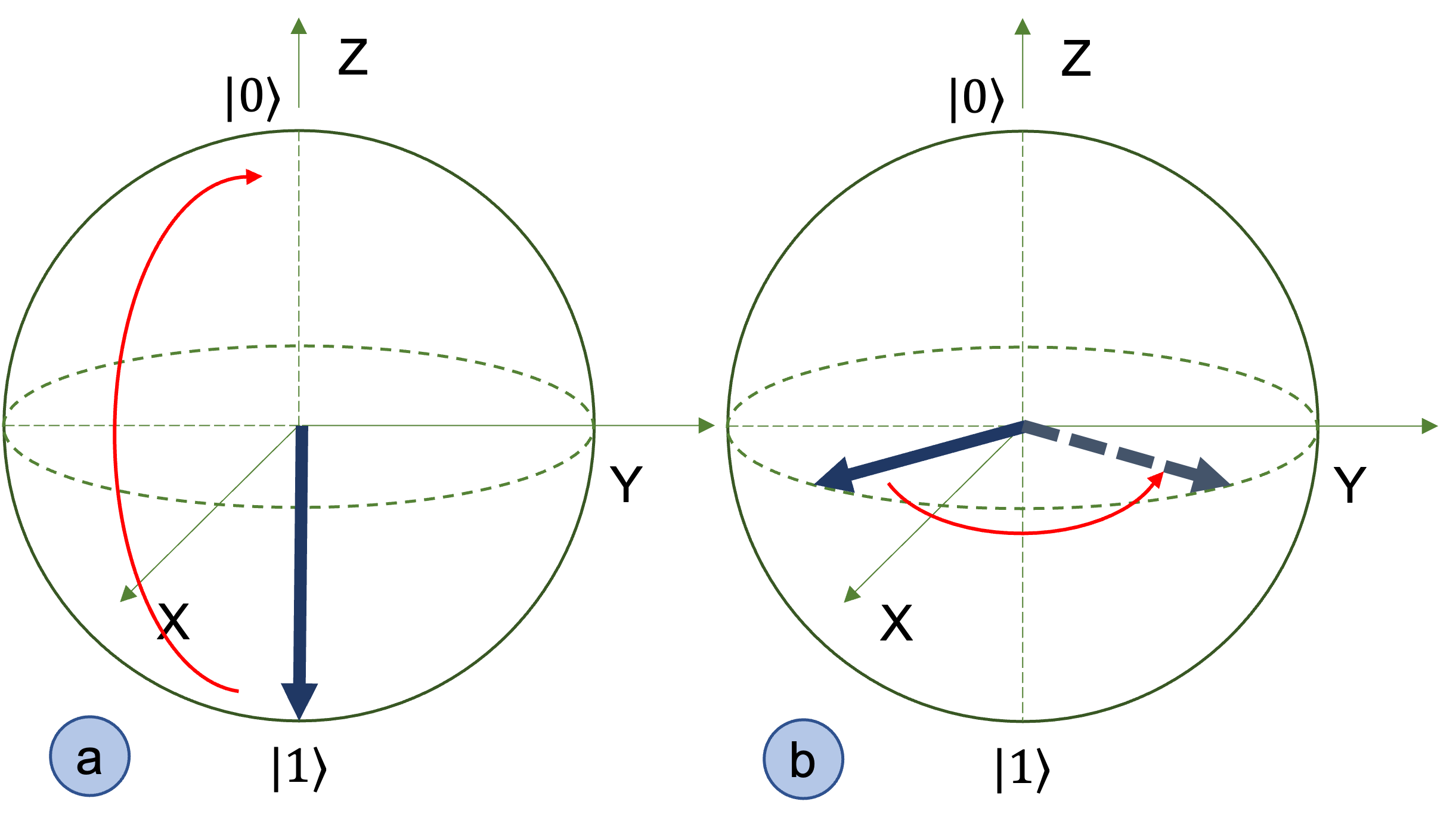}
    \caption{Illustration of a) Relaxation and b) Dephasing of qubit states.}
    \label{2}
\vspace{-4mm}
\end{figure}

Decoherence is related to a short qubit lifetime. Qubits may interact with the environment and spontaneously lose their saved state. For example, Fig. \ref{2} shows the effect of relaxation, one type of decoherence. Due to relaxation, a qubit in state $\ket{1}$ spontaneously loses energy and ends up in state 0. Decoherence of a qubit is usually characterized by T1 relaxation time and T2 dephasing time. If the gate time is tg, then roughly 1--exp(--tg/T1) is the error probability that a state $\ket{1}$ will be damped. This implies that if the gate time (operation) is long, the qubit will lose its state more.
\subsubsection{Measurement Error}
Reading out a qubit containing a state 1 may result in a state 0 and vice versa due to readout error; this arises due to measuring circuitry imperfections. The readout error probability can be quantified using a simple technique. It entails preparing a qubit in all binary states (i.e., 0 and 1 for a single qubit) and reading it out (both preparation and measurement multiple times). The qubits on IBM machines are initially set to 0 states by default. Therefore, to prepare state ‘1’, a quantum-NOT (X) gate has to be applied to the $\ket{0}$ state. Ideally, if we repeat the process of preparing a state 0 or 1 and reading out N times, it should generate 0 or 1 all the time. However, due to readout error, a flipped bit might be read in some of the cases. For example, say state 0 is prepared and measured 1000 times. Out of these 1000 trials, 900 times it reads out 0, and other 100 times it reads out 1. Thus, measurement error rate $M_{01}$ will be 100/1000 = 0.1 ($M_{xy}$ stands for probability of reading out state ‘y’ while the prepared state is ‘x’; thus, $M_{00} = 900/1000 = 0.90$). For multi-qubit readout characterization, the same protocol applies. However, the number of binary states that need to be prepared and read will be higher. For example, to characterize a 3-qubit system 2$^3$ = 8 binary states (000, 001, 010, ..., 110, and 111) need to be prepared and measured (each N-times). Unlike gate-error and decoherence, which depend on the number of gates in the circuit, readout error is gate count agnostic. It solely depends on the state being read. 
\subsubsection{Crosstalk error}
Crosstalk is another kind of error present in the near-term quantum computers. The effect of a gate operation on one qubit should, in theory, be unaffected by what happens on other qubits. Pulses are used to create quantum gates. However, the gate pulse intended for one qubit can accidentally excite an unwanted qubit, which is known as 'crosstalk.' Crosstalk may cause conditional dependence in gate errors. As a result, the gate error of a single gate operation may differ from the gate error of a parallel gate operation. According to \cite{11}, the gate-error with another operation running in parallel can be 2X-3X higher than with an isolated gate operation.

\section{Quantum Hardware} 
\label{sec:hard}
Having introduced quantum computing basics in the prior section this section focuses on the various hardware technologies and developments.
\begin{figure}
    \centering
    \includegraphics[width=4.5in]{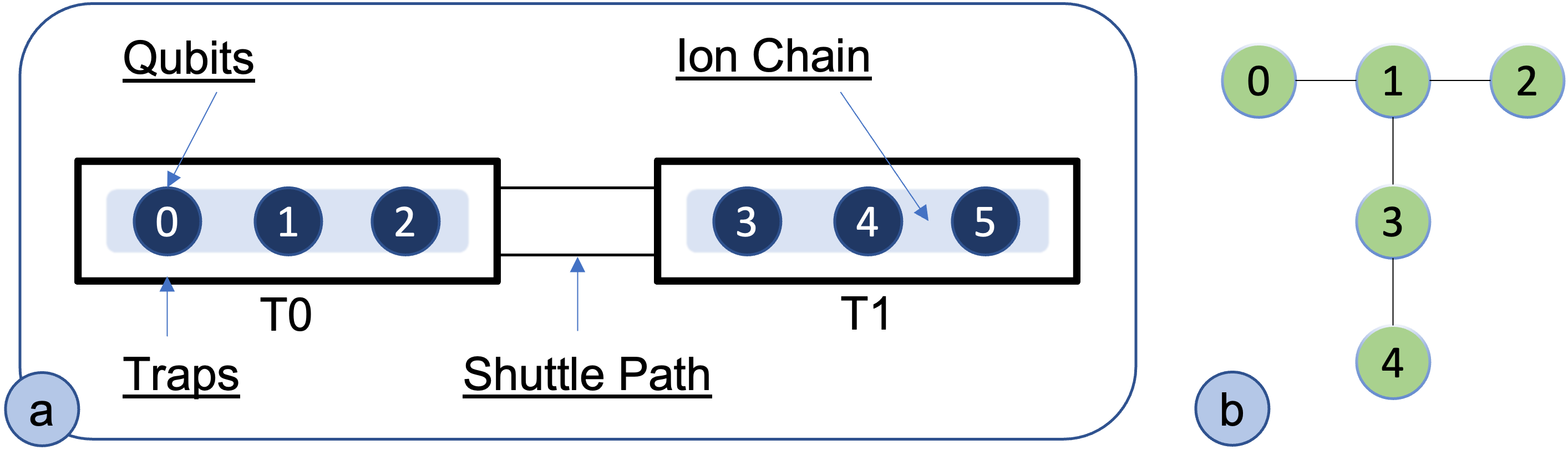}
    \caption{a) A two trap TI system with three qubits each. b) Coupling graph for ibmq$\_$lima superconducting device.}
    \label{3}
\vspace{-4mm}
\end{figure}
\subsection{Qubit Technologies}
A qubit is the fundamental unit of quantum information and the foundation of a quantum computer. At their heart, qubits are two-level systems, which means that every two-level system can physically realize a qubit. There are several technologies such as superconducting, trapped-ion, neutral atom, diamond NV-center, quantum dot, and photon that satisfy the requirement for a qubit. Some of the most common types are:
\begin{figure}
    \centering
    \includegraphics[width=4.5in]{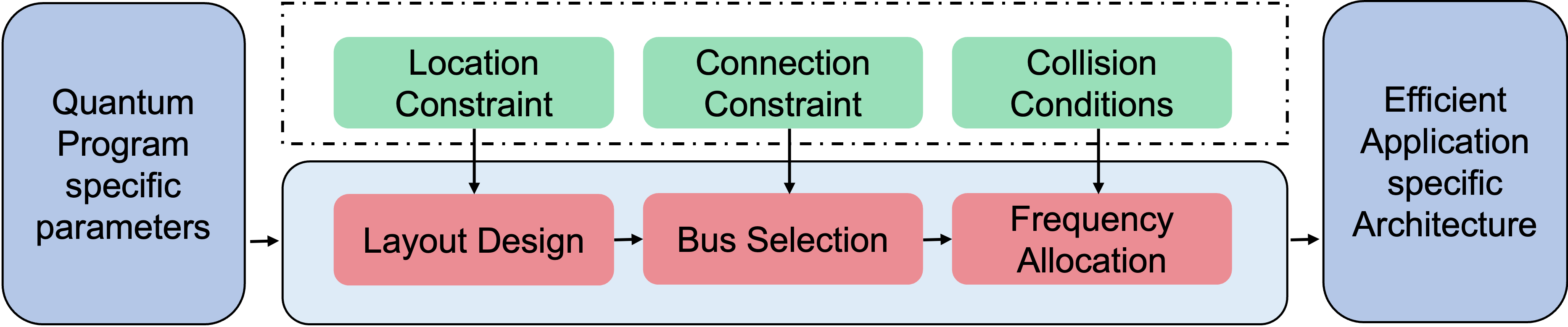}
    \caption{Overview of quantum application specific architecture design flow (adopted from \cite{55}).}
    \label{8}
\vspace{-4mm}
\end{figure}
\subsubsection{Superconducting qubits}
Superconductors allow an electrical current to flow with no resistance when cooled to very low temperatures. We can design electrical circuits based on superconductors to behave like qubits. The idea is to build an anharmonic oscillator. In an anharmonic oscillator, the energy separation between states is different. Therefore, the lowest two energy states are used as a qubit. For harmonic oscillators, the energy states are equally separated, which makes it difficult to control inter-state transition. Superconducting qubits are fabricated by connecting a capacitor and a superconducting Josephson Junction (JJ) in parallel. This assembly works as an anharmonic LC oscillator in which the Josephson Junction works as a non-linear inductor. The JJ requires ultra-low temperature for it to operate in the superconducting regime. Thus, superconducting qubits are usually hosted inside large dilution refrigerators. Ref. \cite{12} gives a comprehensive overview of the current state of play for superconducting qubits. Prominent companies conducting research in superconducting quantum computing \cite{89} are Google, Rigetti, IMEC, BBN Technologies, Intel and IBM\cite{88}.
According to \cite{55}, quantum software and hardware systems should be designed collaboratively in order to fully exploit the potential of quantum computing. They review several architectural design works. One of them is developing a superconducting quantum processor architecture for a specific program \cite{56} in order to achieve a high yield rate with a low mapping overhead. The proposed architecture design flow is depicted in Fig. \ref{8}. They divided the design of a superconducting quantum processor architecture into three key subroutines: layout design, bus selection, and frequency allocation. Each subroutine targets a different hardware component or configuration, incorporating profiling results and physical constraints. They focus on the qubit placement in the layout design and try to make those qubit pairs with more two-qubit gates between them nearby to reduce the mapping overhead. The bus selection subroutine then determines how the physical qubits are linked. According to the profiling information, they only add qubit connections (also known as qubit buses) to the locations that are expected to reduce the mapping overhead the most. Finally, the frequency allocation function will assign frequencies to all physical qubits that have been put. By attempting to eliminate frequency collision scenarios on the created architecture, the subroutine will boost the final yield rate.
\begin{figure}
    \centering
    \includegraphics[width=4.5in]{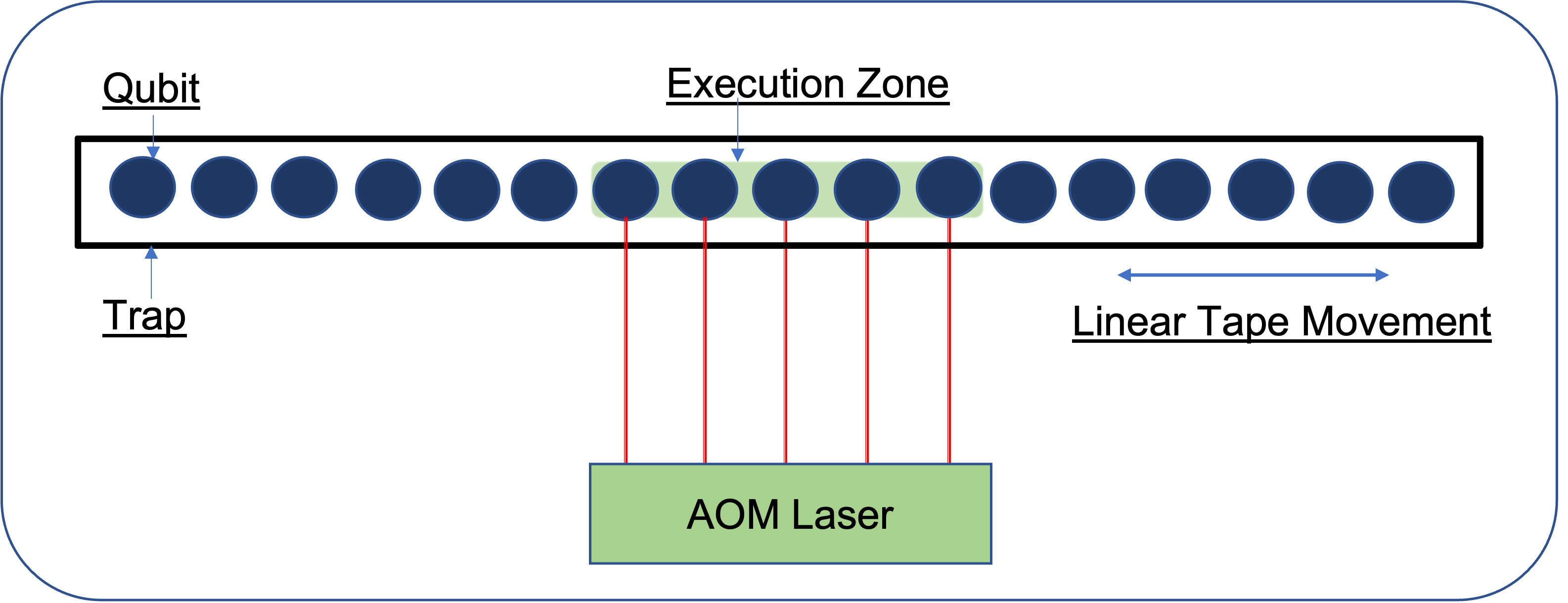}
    \caption{A quantum computer architecture based on trapped ions and linear tapes. Acousto-optic modulators (AOMs) aim laser beams onto ions in the execution zone to perform quantum operations. The entire ion chain is translated until the target qubit is relocated into the execution zone in order to execute gate operations on the other qubits (adopted from \cite{57}).}
    \label{9}
\vspace{-4mm}
\end{figure}

\subsubsection{Trapped-Ion qubits}
Another way of realizing a qubit is by using the energy levels of electrons in neutral atoms or ions. In their natural state, these electrons occupy the lowest possible energy levels. Using lasers, we can “excite” them to a higher energy level and can assign the qubit values based on their energy state. Trapped ion QC system are implemented by trapping ionized atoms like Yb or Ca between electrodes using electromagnetic field \cite{13}. Data $\ket{0}$ and $\ket{1}$ are encoded as internal states such as hyper-fine or Zeeman states of the ions. Qubits are stored in stable electronic states of each ion, and quantum information can be transferred through the collective quantized motion of the ions in a shared trap (interacting through the Coulomb force). Fig. \ref{3}a), illustrates various components of a 2-trap TI system. The ions are organized in form of an ion chain inside a trap. Trap capacity is the maximum number of ions that a trap can accommodate. The traps are connected by a shuttle path which allows movement (shuttle) of an ion from one trap to another if needed. Prominent companies conducting research in Trapped Ion quantum computing are IonQ, Honeywell, Alpine Quantum Technologies and Universal Quantum.
TI systems typically employ a single trap design, which has significant scaling issues. A modular design known as the Quantum Charge Coupled Device (QCCD) has been proposed \cite{39} to advance toward the next significant milestone of 50–100 qubit TI devices. Small traps are coupled by ion shuttling in a QCCD-based TI system. Authors conduct an intensive application-driven architecture analysis to evaluate the major design choices of trap size, communication topology, and operation implementation methodologies in order to realize QCCD-based TI systems with 50–100 qubits. They show that trap sizing and communication topology decisions can effect application dependability by up to three orders of magnitude using several applications as benchmarks and several hardware design points. Another approach to design a hardware architecture for TI systems is discussed in \cite{57}. The authors propose adopting "TILT" (Fig. \ref{9}), a linear "Turing-machine-like" architecture with a multi-laser control "head" in which a linear chain of ions moves back and forth under the laser head, as a building block to extend previous scalable trapped-ion quantum computing approaches. They claim that TILT can significantly decrease communication when compared to Quantum Charge Coupled Device (QCCD) systems of comparable size. The principle behind a TILT design is that operations are only done to ions in the execution zone towards the centre of the trap, and the chain is moved back and forth to allow for long-range interactions. The complex shuttling primitives of a quantum charge-coupled device (QCCD) design are hence not required for such a machine.

\subsubsection{Spin qubits}
Controlling the spin of charge carriers (electrons and electron holes) in semiconductor devices can also be used to implement a Qubit \cite{14}. The majority of quantum particles act like tiny magnets. Spin is the name for this characteristic. The spin orientation is either entirely up or fully down, never halfway up or down. A spin qubit is created by combining these two states. Local depletion of two-dimensional electron vapors in semiconductors such as gallium arsenide, silicon, and germanium has been used to create spin qubits. Some reports also show implementation in graphene \cite{15}.

\section{Quantum Algorithms}
\label{sec:application}

Quantum algorithms are algorithms that run on a realistic model of quantum computation, (the most used one being the quantum circuit model \cite{16}\cite{17}) and that are inherently quantum or use some essential feature of quantum computation such as quantum superposition or quantum entanglement. Problems that are fundamentally unsolvable by classical algorithms \cite{18} cannot be solved by quantum algorithms either. However, the added value of quantum algorithms is that they can solve some problems significantly faster than classical algorithms as they leverage quantum phenomenon such as superposition, entanglement and interference. Applications of quantum computing are as diverse as the fields necessary to create quantum information processing technology. There is an extensive literature on quantum algorithms that has been developed \cite{19}\cite{20}. The field is now entering the era of noisy intermediate-scale quantum (NISQ) devices \cite{21}(quantum computers that are sufficiently large-tens to hundreds or a few thousand qubits that they cannot be efficiently simulated by a classical computer but are not fault tolerant). Noise was only examined formally and proved to be theoretically surmountable in the early stages of quantum computation, with considerable involvement from the mathematics and computer science communities. As a result, the first wave of quantum algorithms considered that quantum devices would operate without making any noise (or otherwise fully quantum error corrected systems). Since the introduction of NISQ devices, a second wave of quantum algorithms has sprung out, taking into account noise and breakthroughs in algorithm design on traditional computers.. In this section we aim to provide a brief insight into the literature. This section is divided into two subsections: Algorithms designed for Fault tolerant Quantum computers and Algorithms that run on NISQ computers.

\subsection{Algorithms designed for Fault tolerant Quantum computers}
The initial quantum computing algorithms were developed with an ideal quantum computer in mind, with the quantum gate model studied largely without noise. \cite{22} is the canonical reference for this wave of quantum algorithm development, and it remains a reliable reference for the theoretical basis of quantum computing and quantum information to this day. The best-known algorithms are Shor's algorithm for factoring and Grover's algorithm for searching an unstructured database or an unordered list.
\subsubsection{Shor’s Algorithm}
Shor’s algorithms describe two quantum algorithms for integer factoring and discrete logarithm exponentially faster than the best-known classical algorithms \cite{23}.Because of the apparent speedup compared to classical algorithms and the implications of this speedup for known applications, it is a notable and celebrated scientific contribution to quantum computing. Shor's algorithms take advantage of both quantum parallelism and entanglement. There are two sections to the algorithm. The first portion of the algorithm converts the factoring problem into a problem of determining a function's period and can be implemented in a traditional way. The quantum speedup is determined by the second portion, which uses the quantum Fourier transform to find the period. Essentially, the paper \cite{23} shows that the factoring problem is equivalent to the problem of finding the period in a sequence of numbers, although a sequence of numbers that is exponentially longer than the number of bits of the corresponding number to be factored. Thus, while this equivalency does not provide any help in solving the problem on a classical computer (since it would need to generate this sequence of 2n numbers for an n-bit number to factor, which would take an exponential amount of time), it is a perfect problem for a quantum computer as it can be encoded into merely n qubits, and generated in a time that is polynomial in n. Once that sequence is generated, the QFT can be used to find the period. Shor's method, if implemented on a perfect quantum computer, would allow the secret key of the most frequently used public key cryptosystem, RSA, to be computed, meaning that public key encryption might be readily broken.
\subsubsection{Grover’s Algorithm}
Grover’s Algorithm also known as the quantum search algorithm was introduced by Lov Grover in 1996 \cite{24}. It is used for searching an unsorted database with N entries in O($\sqrt{N}$) time and using O(logN) storage space. Searching an unsorted database traditionally involves a linear search, which takes O(N) time. Grover's technique, on the other hand, takes O($\sqrt{N}$) time and is the fastest quantum algorithm for doing so. Unlike other quantum algorithms, which can provide exponential speedup over their classical equivalents, it delivers a quadratic speedup. When N is big, even quadratic speedup is significant. It may also be used to calculate the mean and median of a group of values, as well as to solve the collision problem. It can also be used to tackle NP-complete problems by doing exhaustive searches across all feasible solutions. This would result in a significant speedup as compared to traditional techniques. Grover's algorithm can also be applied to speed up broad classes of algorithms.\cite{25} Grover's algorithm is probabilistic like all quantum algorithms, in the sense that it gives the correct answer with high probability. The probability of failure can be decreased by repeating the algorithm.
\subsection{Algorithms for NISQ computers.}
The near-term quantum devices have limited number of qubits. Moreover, they suffer from various types of noises (decoherence, gate errors, measurement errors, crosstalk, etc). Due to these constraints, these machines are not yet fully capable of executing quantum algorithms requiring high orders of error correction (such as Shor's factorization or Grover's search). However, algorithms such as Quantum Approximate Optimization Algorithm or QAOA, Variational promises to achieve quantum advantage with near-term machines because they are based on a variational principle that does not necessitate error correction \cite{26}. Most of these approaches utilize a conventional computer to perform an optimization procedure using information extracted from the quantum device, usually in an iterative fashion. These quantum optimization methods have been applied to diverse areas such as quantum machine learning \cite{27}.
\subsubsection{Variational Quantum Eigensolver or VQE}
The variational quantum eigensolver (or VQE), was introduced by Peruzzo et el \cite{28}. The basic concept which is at the heart of VQE is that the computed energy of the ground (lowest energy) state of a quantum chemical system decreases as the approximations to the solution improve, asymptotically approaching the true value from above. The input is a rough estimate at the solution, and the output is a slightly improved approximation. This output is then utilized as a guess for the next iteration, and the output grows closer to the correct solution with each cycle. The problem is split down into a series of smaller problems that can be estimated independently in VQE, with the sum of all outputs corresponding to the approximate solution of interest. The process is repeated until a heuristic stopping criteria is met, which is usually equivalent to reaching an energy threshold.
\subsubsection{Quantum Approximate Optimization Algorithm or QAOA}
The Quantum Approximate Optimization Algorithm (QAOA) is a hybrid quantum-classical variational algorithm designed to tackle combinatorial optimization problems \cite{29}\cite{30}. To optimize an objective function, it employs classical optimization of quantum operations. The algorithm is similar to the VQE algorithm in that it starts with a series of preparation and measurement experiments before being optimized by a traditional computer. When sampled, the resulting quantum state gives approximate or exact answers to the computational task.
\begin{figure}
    \centering
    \includegraphics[width=4.5in]{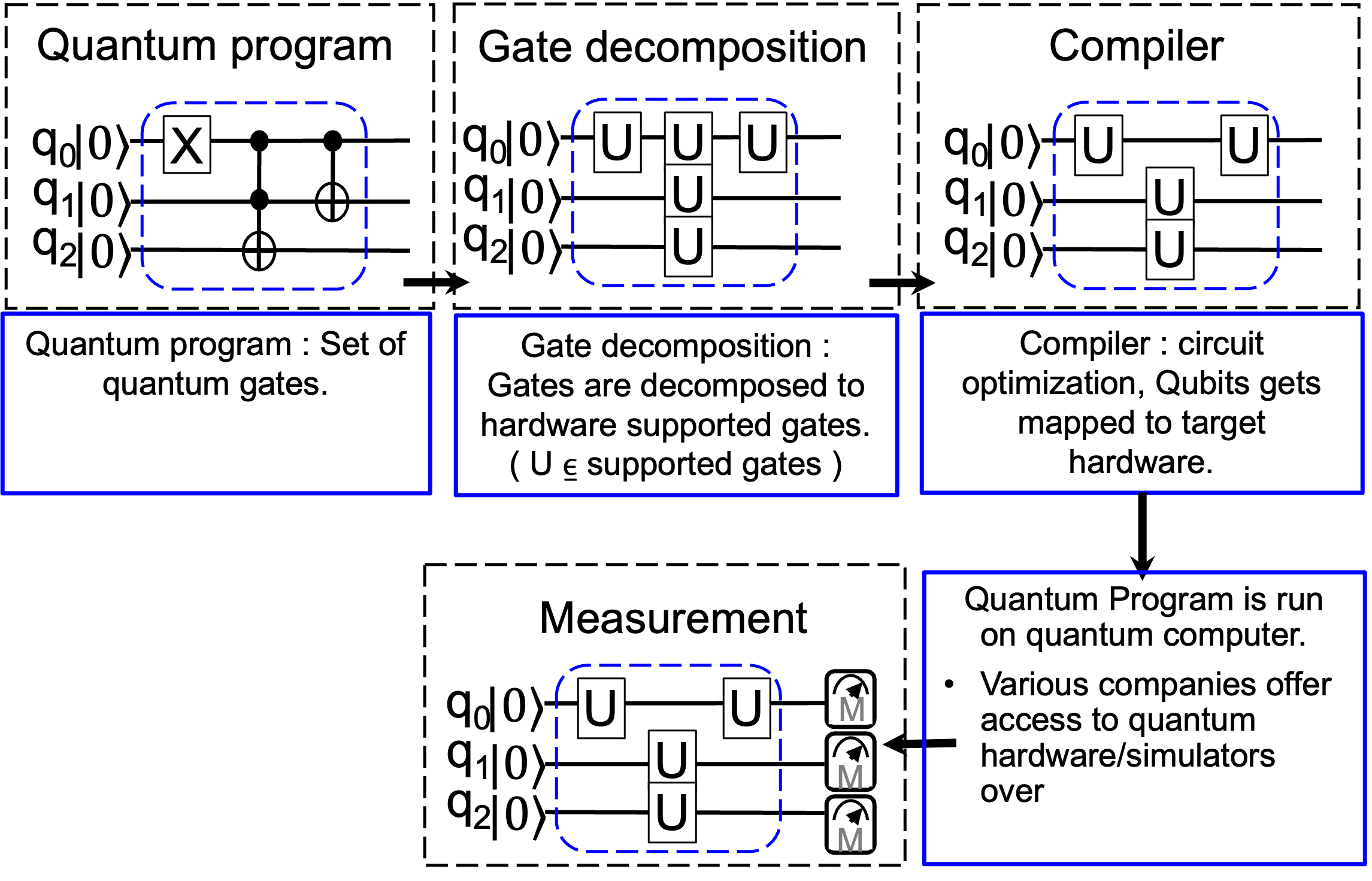}
    \caption{Overview of program flow for quantum computers.}
    \label{4}
\vspace{-4mm}
\end{figure}

\section{Quantum Software}
\label{sec:soft}

A problem is first established, then translated/optimized by software support so that it may be solved efficiently by hardware or simulators in any computing ecosystem. A software suite typically contains programming languages, compilers for mapping algorithms to machines, simulation, debugging, and optimization tools to aid in the efficient implementation of algorithms on systems. Simulation tools, in particular for quantum computers, can allow a programmer to model each quantum operation and track the quantum state that arises, as well as its progress over time. Debugging both applications and newly built hardware need this capabilities. Resource estimators and other optimization tools would allow for quick estimation of the performance and qubit resources required to run various quantum algorithms. This enables a compiler to transform the desired computation into an efficient form, minimizing the number of qubits or qubit operations required for the hardware in question. The optimization heuristics will also depend on the type of hardware the quantum program is going to be run on. Each hardware presents unique set of challenges. In this section we review compilation, mapping and optimization heuristics used for two of the most common quantum hardware – Superconducting Quantum computers and Trapped Ion quantum computers.

\subsection{Quantum Program, Quantum Instruction Sets and Software development kits.}
\begin{figure}
    \centering
    \includegraphics[width=3.2in]{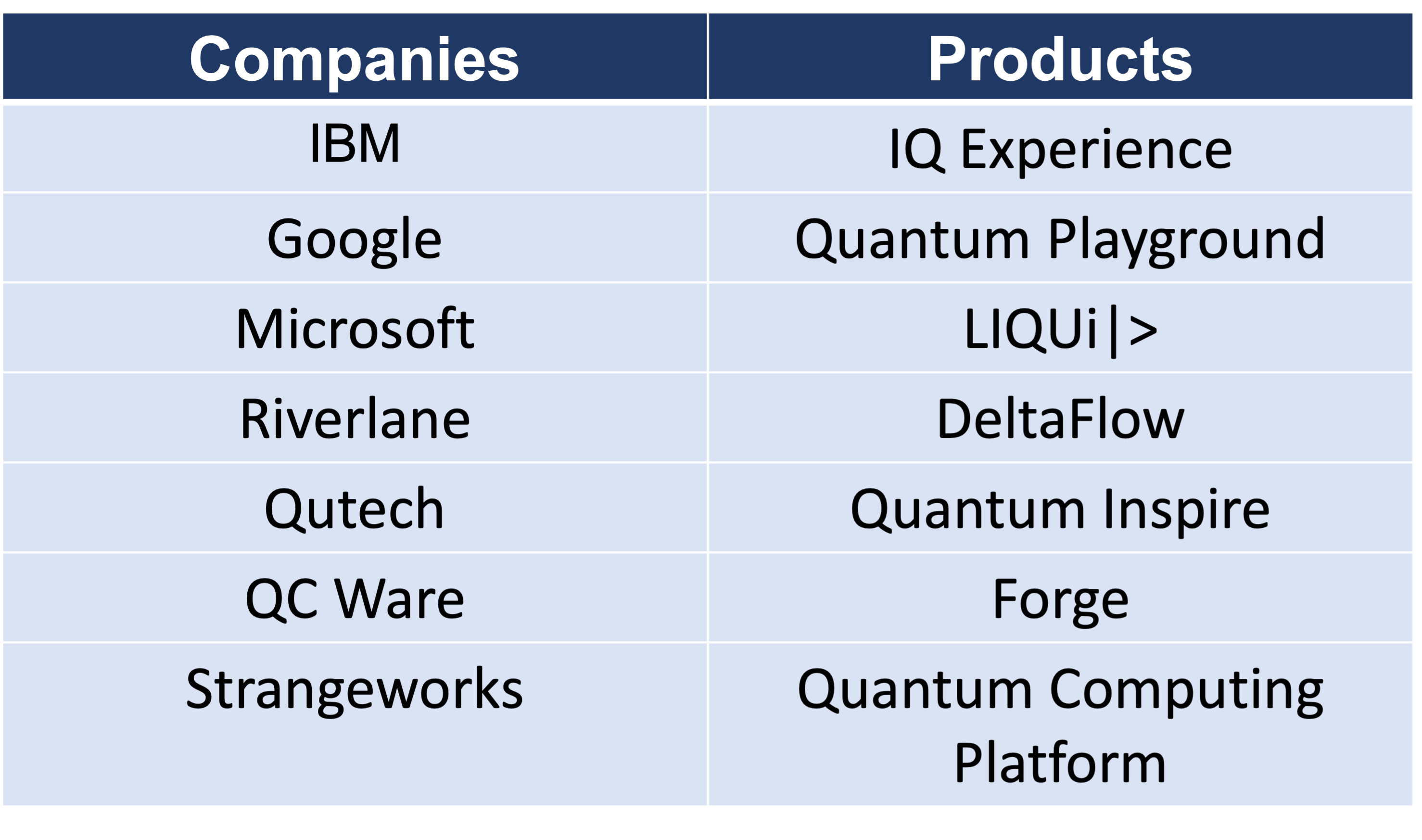}
    \caption{Various leading companies in quantum domain and their products.}
    \label{5}
\vspace{-4mm}
\end{figure}

A quantum program can be represented in the well adopted quantum circuit model \cite{16}. Fig. \ref{4} illustrates brief overview of a quantum program ecosystem. It can be modeled as a series of quantum gates operating on qubits to converge the output to a given solution, similar to classical computing. A quantum program is made up of variables that are logical qubits and quantum operations that can change the state of the qubits. Quantum instruction sets are used to turn higher level algorithms into physical instructions that can be executed on quantum processors. There are various instruction set architectures available such as cQASM, OpenQASM, Quil, Blackbird etc. Quantum software development kits are bundles of tools that allow users to construct and manipulate quantum programs. The users are provided access to quantum hardware where they can simulate the quantum programs or prepare them to be run using cloud-based quantum devices. Fig. \ref{5} illustrates the companies and their products that are the leading computational platforms for quantum computing.

\subsection{Quantum programming languages}

The work published by Bettina Heim and group provides an overview of Q\#, Qiskit, Cirq, Quipper and Scaffold as well as the tools/ecosystems that surround them, and how they have served as a foundation for current and future work\cite{62}. \textbf{Q\#} is a hardware-agnostic quantum programming language designed to enable the execution of large-scale applications on future quantum hardware \cite{63}. As a result, rather than following the imperative style encouraged by assembly-like languages, Q\# focuses on providing high-level abstractions that facilitate reasoning about the intended functionality. It is notable for its support for expressing arbitrary classical control flow. This is in contrast to other quantum programming languages, where this capability is frequently provided by a classical host language. Unlike other quantum programming languages geared toward formal verification, qubits in Q\# are treated like any other data type. The associated libraries, the Q\# compiler, and all other components of the quantum development kit are open source. \textbf{OpenQASM} is a quantum program intermediate representation based on gates\cite{64}. It expresses quantum programs as lists of instructions, which are frequently intended to be consumed directly by a quantum processor. OpenQASM supports abstractions in the form of quantum gates, which can be built in a hierarchical fashion using a set of intrinsic primitives assumed to be available on the targeted processor; for example, a Toffoli gate made up of CNOT gates, T gates, and H gates. In addition, OpenQASM supports single-qubit measurement and basic classical control operations. Qiskit provides a Python-based programming environment for creating and manipulating OpenQASM programs \cite{65}. It includes extensive simulation capabilities, such as state vector and density matrix simulators that can be run on both CPUs and GPUs, in addition to support for execution on quantum processors. As a result, it allows users to simulate the effects of noise defined by any custom model, including arbitrary Kraus operators. The online documentation \cite{66}, which includes tutorials and is generated for each release, provides a good overview of the full range of capabilities included in Qiskit. \textbf{Cirq} is a Python quantum programming library that focuses on supporting near-term quantum hardware. Cirq's primary goal is to enable the development of quantum programs capable of running on quantum computers available now or in the near future that lack error correction (NISQ hardware) and are subject to certain device topologies. It includes mechanisms for fine-tuning how a quantum program executes on the specified quantum hardware, as well as tools for simulating hardware constraints such as, noise limitations or the physical layout of the qubits\cite{67}. In contrast to other languages where qubits can be allocated dynamically, layout in Cirq is done manually. It is built into Python. Python's control flow constructs, such as if and while test statements, can be used to build a circuit before execution. Cirq includes device models for many of Google's quantum processors\cite{67}, such as Bristlecone and Sycamore.

\textbf{Quipper} is a circuit description language, which means it can be used to construct circuits by applying gates on qubits in an organized manner. The circuits themselves are data that can be provided to functions in the host language Haskell for circuit optimization, resource estimation, or error correction, for example. Prototypical implementations of Quipper-like languages, such as Proto-Quipper-S\cite{68}, Proto-Quipper-M\cite{69}, and Proto-Quipper-D\cite{70}, have evolved with the purpose of enforcing quantum-specific features such as the quantum information no-cloning theorem. \textbf{Scaffold} is a stand-alone programming language. It is intended to be similar to existing traditional programming languages, like C: Scaffold uses the imperative programming model of C, as well as many of its recognizable features such as functions (called modules in Scaffold), if statements, loops, structures, and pre-processor directives\cite{71}. Scaffold programs can also automatically convert conventional functions into reversible logic, which is done using quantum gates, and then incorporate it as an oracle in a larger quantum algorithm\cite{71}. \textbf{Intel Quantum SDK} Intel recently demonstrated it's Quantum SDK at IEEE Quantum Week, held in Colorado, USA, 2022 \cite{90}. It provides developers with tools to help them learn how to program quantum algorithms.  It is based upon the C++ programming language and uses the LLVM intermediate level description from classical computing as a base. It is designed to work with hybrid classical/quantum variational algorithms and will be compatible with other components of Intel's quantum stack, such as high-performance quantum simulators and, eventually, Intel's spin-qubit-based quantum processor. The beta version is accessible via the Intel Developer Cloud.

Other quantum programming languages and open-source software frameworks include Forest/PyQuil\cite{72}, ProjectQ\cite{73}, QWIRE\cite{74}, staq\cite{75}, Strawberry Fields\cite{76}, tket\cite{77}, XACC\cite{78}, and QuTiP\cite{79}.

\subsection{Quantum Annealing}

There are various approaches to building quantum computing hardware, such as, universal gate model quantum computers or quantum annealers. Universal gate model quantum computing, also known as general purpose quantum computing, is the most powerful and flexible type of quantum computer, but it is difficult to build and maintain the stability of qubits. It is based on creating quantum circuits with stable qubits and using them to solve problems. However, maintaining stability for qubits is difficult. As the number of qubits increases, so does the complexity of the problem. Quantum annealing, on the other hand, is less affected by noise than gate model quantum computing and focuses on the solution of NP Hard problems. This feature enables greater qubit usage and thus more parameters for specific problems. Quantum annealing (QA) is an optimization approach that uses quantum fluctuations to discover the global minimum of a given objective function over a given set of candidate solutions (candidate states). It is mostly employed for issues where the search space is discontinuous (combinatorial optimization problems) with multiple local minima, such as determining the ground state of a spin glass or solving the traveling salesman problem. \cite{80} B. Apolloni, N. Cesa Bianchi, and D. De Falco coined the phrase ``quantum annealing" to describe a quantum-inspired classical method in 1988. Quantum annealing can be compared to simulated annealing, in which the "temperature" parameter functions similarly to the tunneling field strength in QA. The temperature determines the likelihood of shifting to a higher ``energy" level from a single current state in simulated annealing. The quantum-mechanical probability of changing the amplitudes of all states in parallel is determined by the strength of the transverse field in quantum annealing. Under some situations, analytical \cite{81} and numerical \cite{82} evidence suggests that quantum annealing beats simulated annealing \cite{83}.

Quantum annealing begins with a quantum-mechanical superposition of all potential states with equal weights. The system then evolves in accordance with the time-dependent Schrödinger equation, which is a natural quantum-mechanical evolution of physical systems. According to the time-dependent strength of the transverse field, which induces quantum tunneling between states, the amplitudes of all candidate states change, resulting in quantum parallelism. If the rate of change of the transverse field is slow enough, the system remains near to the instantaneous Hamiltonian's ground state. If the rate of change of the transverse field is increased, the system may briefly leave the ground state but has a higher possibility of reaching the final problem Hamiltonian's ground state, i.e., diabatic quantum computation. \cite{84} Finally, the transverse field is turned off, and the system is assumed to have arrived to the ground state of the classical Ising model, which corresponds to the solution to the original optimization issue. Immediately following the initial theoretical idea, \cite{85} an experimental proof of the success of quantum annealing for random magnets was reported. T. Kadowaki and H. Nishimori developed Quantum Anneling in its current form in ``Quantum annealing in the transverse Ising model". D-wave was the first to use a quantum annealing approach to build a quantum computer. D-Wave Systems announced the first commercial quantum annealer on the market, the D-Wave One, in 2011. This system employs a 128-qubit processor chipset \cite{86}.

\subsection{Compilation, Mapping and Optimization}

Quantum compilation bridges the gap between the computing layer of high-level quantum algorithms and the layer of physical qubits with their specific properties and constraints. Quantum circuit optimization is an essential component of the quantum computing toolchain. Many Noisy Intermediate-Scale Quantum (NISQ) devices maintain only loose connectivity between qubits, which means that a valid quantum circuit frequently requires swapping physical qubits to satisfy adjacency requirements. Optimizing circuits to reduce such swaps and other parameters is critical for using quantum hardware in the near future. A significant family of optimal synthesis algorithms function by completely enumerating all circuits and returning the lowest cost circuit that can do the specified computation; this technique is known as exhaustive or brute-force searching. This method is quite popular in the circuit synthesis community for optimally assembling small frequently used gates or functions to the target gate set, and it can be very effective in these modest instances. Shende et al. \cite{58} synthesized all minimal gate count circuits for reversible functions on 3 bits using a breadth-first search over the gate set {X, CNOT, T OF}. While breadth-first searches are prevalent in reversible circuit synthesis, the lack of efficient unitary representations complicates such approaches. Fowler \cite{59} avoided this issue by conducting the breadth-first search directly, that is, without the assistance of a pre-computed database with efficient lookup. Non-search based synthesis has been utilized in quantum computing on occasion. Kliuchnikov et al. \cite{60} in particular provide an approach for decomposing an arbitrary single-qubit unitary. The earlier described algorithms were largely concerned with lowering gate counts, and any depth reduction was a byproduct of that. However, when there are many computational resources available, it can frequently make sense to raise complexity in order to parallelize operations to take advantage of the extra resources, as in classical computing. However, when there are many computational resources available, it can frequently make sense to raise complexity in order to parallelize operations to take advantage of the extra resources, as in classical computing. Broadbent and Kashefi \cite{61} develop an algorithm for translating quantum circuits to a pattern (a computation in the measurement-based model) that adds a number of additional ancillas linear in the number of gates. Mapping refers to assigning logical qubits to the physical qubits of the hardware. 

In the following section we discuss compilation, mapping and optimization specifically for Superconducting and Trapped Ion quantum computers.

\subsection{Superconducting Quantum Computers}

\subsubsection{Coupling Constraints and need for SWAP operation}
A qubit in superconducting quantum systems is connected to one or more neighboring qubits using resonators (waveguides) that allow a multi-qubit gate between them. Fig. \ref{3}b) depicts the qubit connectivity graphs for an IBM computer (IBM$\_$lima). The nodes (the circles) represents the qubits. Through coupling graph, it is understood that the native 2-qubit gate (CNOT in IBM and CZ in Rigetti) can only be applied between connected qubits. For instance, CNOT between qubit-1 and 2 is allowed on IBM$\_$lima device as there exists an edge between these qubits in the graph. However, CNOT cannot be applied directly between qubit-1 and 3 as they are not connected. This limited connectivity presents a challenge in quantum circuit mapping and is often referred to as coupling constraints. The constraint is handled by routing qubits via the SWAP operation so that logical qubits with 2-qubit operations become nearest neighbors. Qubit Mapping is the term used in the literature to describe the process of changing a quantum circuit to fit hardware restrictions. The final SWAP-inserted version is a nearest-neighbor (NN) compliant circuit that can be run on a quantum computer directly. It's worth noting that the additional SWAP operations must be decomposed to the target hardware's basic gates before being executed. Any traditional Qubit Mapping procedure entails a)selecting physical qubits on the hardware for the logical qubits in the circuit (qubit allocation), b)initial one-to-one mapping of the logical and physical qubits (initial placement), and c)adding (as few as possible) SWAP operations to meet the hardware constraints for the entire circuit.

\subsubsection{Compilation and Optimization}
Additional SWAP operations to fulfill the communication need among qubits is an NP-complete problem, according to \cite{31}. In the literature, there are two different techniques to reducing the amount of SWAP operations during the qubit mapping procedure. In the first approach, the problem is formulated as an instance of constraint satisfaction problem, and later, powerful reasoning engines (e.g., SMT solver, ILP solver, and SAT solver) are used to find the best possible solution to meet these constraints \cite{32}\cite{33}\cite{34}. Although such approaches frequently yield near-optimal results for small circuits, they are unsuitable for big circuits, resulting in significant compilation time overhead. 
The second method is based on effective heuristics that gradually lead to a solution \cite{35}\cite{36}. However, decisions at each stage are made with the goal of maximizing the gain in the current step, which may result in sub-optimal results (e.g., local optima). The Qubit Mapping problem is solved using A* heuristics \cite{37}. The proposed approach chooses a single SWAP operation that minimizes the cumulative SWAP distances of all the two-qubit operations in the current layer in each algorithm iteration. The minimum number of SWAP operations required to produce a logical qubit nearest neighbor to another qubit is known as SWAP distance. The technique is repeated until the cumulative SWAP distance for the current layer reaches zero. The algorithm moves on to the next layer after finding a set of SWAP operations that discovers a new logical-to-physical qubit mapping that meets the hardware restrictions to execute all of the gate operations in the current layer.The look-ahead strategy is a variation of their primary approach that included the cumulative SWAP distances of the following layers in the cost function to determine SWAP operations for the present layer. The look-ahead method frequently yields a superior solution at the cost of increased compilation time.
The approach in \cite{37} verifies all the mapped qubits as well as the qubits related to them before adding a SWAP gate. However, it has been reasoned \cite{36} that we can minimize this number because not all physical qubits in the routing choice have the same 'priority.' Priority qubits are active qubits in the 'front layer' and the qubits related to them. They use a lesser number of SWAPs to examine the SWAP between these reduced sets of qubits for routing. For bigger quantum computer designs and quantum circuits, this reduction can be significant. \cite{38} demonstrates another routing optimization aspect. The control and target qubits of the CNOT under discussion are reserved in the proposal, and they use one-bends paths for routing. 
\begin{figure}
    \centering
    \includegraphics[width=3.2in]{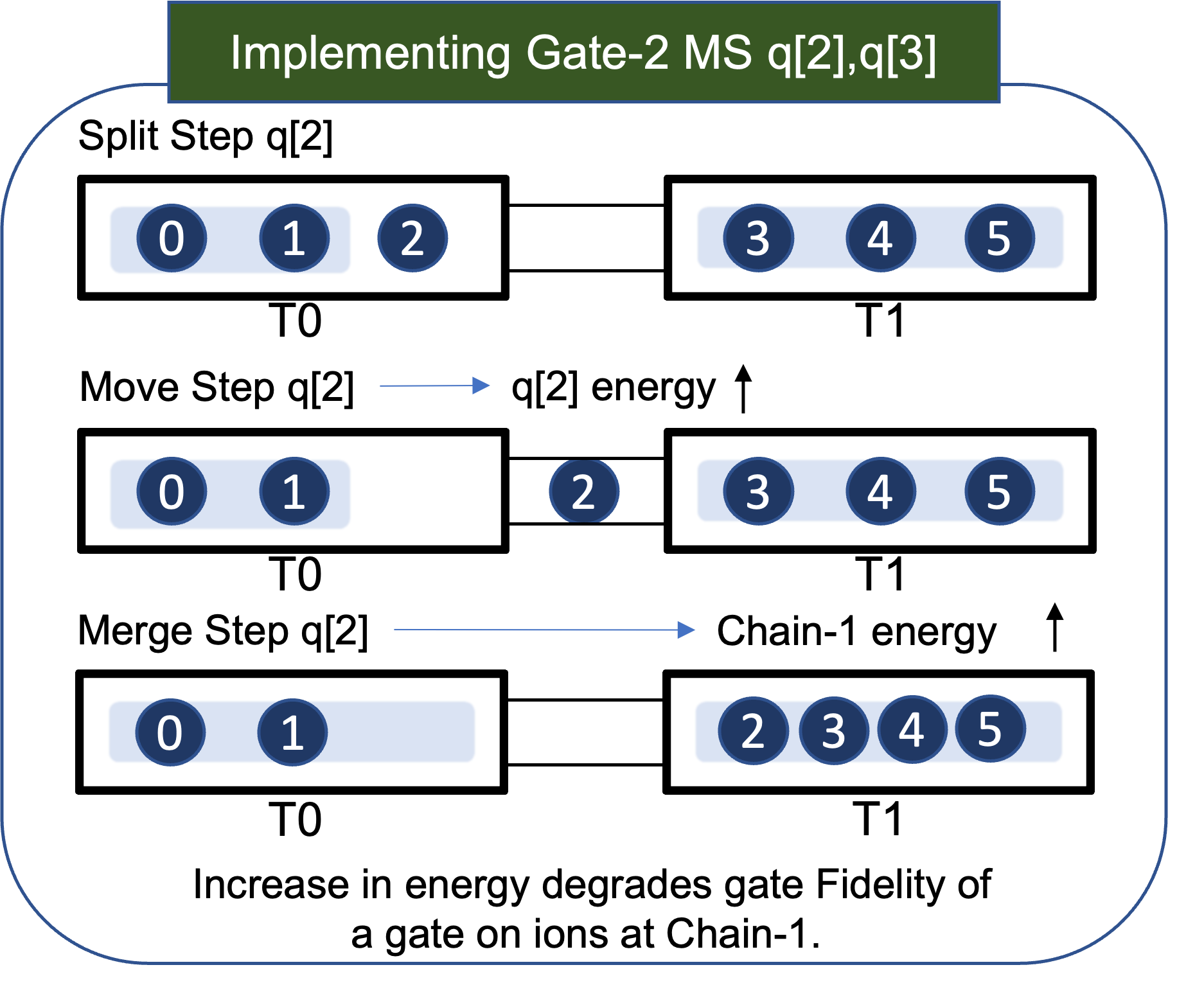}
    \caption{Shuttle steps to move ion-2 from trap T0 to trap T1.}
    \label{6}
\vspace{-4mm}
\end{figure}

\subsection{Trapped-Ion Quantum Computers}

\subsubsection{Shuttle Operation}

A major hurdle in realizing large TI systems is confining many ions in a single trap as it decreases the spacing between ions, making it challenging to pulse a qubit using laser controllers selectively. Moreover, the gate time becomes slow, which results in longer program execution time. Therefore, the pathway to scalability in TI systems involves multiple interconnected traps. However, in a multi-trap system, computation is sometimes required on data from ions situated in different traps. For such cases, one ion needs to be shuttled (moved) from one trap to another so that the ions are co-located and the gate operation can be performed. A compiler adds shuttle operations to a quantum program to satisfy the inter-trap communication, however, the shuttle operation increases program execution time and degrades quantum gate fidelity (The gate fidelity (F) is usually defined as the complement of the error rate. A lower gate fidelity will introduce more errors in the output and can completely decimate the result). The shuttle operation involves several steps as shown in Fig. \ref{6}.For example to implement gate MS q[0], q[1], it involves ions from the same trap (T0) and can be executed directly. However, to execute gate MS q[2], q[3] involves ions from different traps. Therefore, a shuttle operation is needed to bring both ions into the same trap. For the shuttle operation, first, ion-2 is split from Chain–0 and shuttled from T0 to T1, adding energy to the ion. Then, ion-2 is merged to the Chain-1. Finally, gate MS q[2],q[3] can be executed as the ions are in the same trap (T1). This increase in energy degrades gate fidelity. Therefore, it is essential to minimize the number of shuttle operations.

\subsubsection{Compilation and Optimization}
Murali et al. \cite{39} proposed the first compiler for a multi-trap TI system. They studied trap geometry, trap size, and gate implementation methods in depth. Several compilation policies were also offered, including an initial mapping policy, a shuttle direction policy, and a traffic block resolution policy. Realistic hardware performance parameters are included in the toolchain. It compiles an application (quantum circuit) to address communication needs by adding more shuttle operations. The architectural policies aim to keep the number of shuttles to a minimum. The tool creates a machine-executable circuit and reports on its dependability, operation count, and execution time.
Another recent study on the compilation for TI systems is reported in Ref. \cite{40}. They created a compiler for the linear tape model, a TI technology variation. A chain of ions is moved back and forth around a gate zone in the linear tape model to apply gate excitation to different qubits. Because back-and-forth movement of the ion chain increases chain energy, lowers gate quality, and increases execution time, the compiler strives to minimize it. Abdullah Ash Saki et el in \cite{41} presents compiler optimizations for multi-trap TI quantum computers. 
Furthermore, the improved compilation enhances the program fidelity up to 22.68X with a modest increase in the compilation shuttles compared to previous state-of-the-art.

\section{Considerations for Noisy systems}
\label{sec:noisy}
In quantum systems single-qubit, multi-qubit gate errors, relaxation and dephasing times, and measurement errors are prominent sources of noise. These errors have varying error rates, implying that some qubits or qubit couplings are less erroneous than others. As a result, noise awareness can be employed in the software stack to optimize a program for a certain hardware to improve program execution dependability. In this section, we look into noise resiliency research that is both technology neutral and Superconducting technology specific.
\subsection{Technology Agnostic Work}

\subsubsection{Noise-aware Qubit Mapping}
Some qubits are better (less erroneous) than others at performing computations. Several noise-aware mapping algorithms have been developed as a result of this observation \cite{42}\cite{43}\cite{44}. Prioritizing less erroneous qubits to conduct the majority of gate operations being the key strategy. The main approach being prioritizing less erroneous qubits to perform the majority of the gate operations.
The authors of \cite{43} employed the satisfiability-modulo-theorem (SMT) to make decisions about qubit allocation and movement while accounting for error rate changes. In addition to gate error, they factored readout errors into their allocation choice. Their weighted technique gives users the freedom to choose between gate and measurement errors. \cite{44} proposes policies such as variation-aware qubit allocation (VQA) and variation-aware qubit mobility (VQM). To increase the program's success rate, the authors propose harnessing qubit-to-qubit variance. VQA selects a set of physical qubits with the highest cumulative connectivity strength. The cumulative coupling strength represents two things: (a) a qubit is coupled to more neighbors, which is favorable for optimum routing (less SWAP), and (b) the 2-qubit operations between the qubit and its neighbors will be less erroneous. Furthermore, the VQM policy ensures that the compiler choose a routing path with fewer erroneous links. In \cite{42}, the authors present QURE to schedule gate operation to less noisy qubits intelligently and thus, resulting in better fidelity of the output state. They propose two approaches, (a) isomorphic sub-graph (ISG) search and (b) greedy, to find a better allocation of program (logical) qubits to hardware (physical) qubits. They propose using the ISG search approach to start with an optimal depth version of a quantum circuit and check multiple isomorphic sub-graphs systematically. Each sub-graph is given an approximate success probability, and the sub-graph with the highest success probability is chosen to execute the circuit. They demonstrated that QURE can improve correct output probability or fidelity by a large margin without incurring any physical or circuit level overhead in a rigorous simulation using a model noisy quantum system and an experiment with IBM's real quantum device.

\subsubsection{Measurement Error Mitigation}
The final state of a quantum circuit is measured once it has completed its execution. However, due to readout error, reading out a qubit containing a 1 may result in a 0 and vice versa; this arises due to measurement circuitry imperfections. A single-qubit measurement, for example, is performed in two steps in IBM's superconducting quantum hardware: (a) executing the readout pulse on the qubit's readout channel, and (b) recording the associated signal, which measures the qubit's energy state, on the acquisition channel. The signal collected during the course of the acquisition is summed to produce a single complex value, which is then plotted in an I-Q plane, with $\ket{0}$ and $\ket{1}$ meant to represent separate clusters. To classify the measured state ($\ket{0}$ or $\ket{1}$) from the imaginary and real components of the complex value, IBM currently utilizes a linear classifier. A synthetic dataset is used to train the classifier. To construct this dataset, the qubit is prepared in the $\ket{0}$ and $\ket{1}$ states many times, followed by measurement operations. The input features to the classifier are the real and imaginary components, and the actual states are the labels. The authors of \cite{45} demonstrated that the linear classifier has non-uniform measurement errors. When the true state is closer to $\ket{1}$, the error magnifies significantly. The considerable overlap zone generated by the linear decision boundary between the $\ket{1}$ and $\ket{1}$ states contributes to the error magnification. The authors presented two non-linear classifiers (based on circular and elliptical decision boundaries) and trained them to minimize the variance in measurement errors across different states to get around this problem. The variation of the errors was reduced significantly over the linear classifier measurement, according to the authors.

\subsection{Superconducting-specific work}

\subsubsection{Crosstalk Mitigation}
As two gates run in parallel, the crosstalk error occurs, resulting in an increase in gate faults of two parallel gates when compared to isolated gates. The authors of \cite{46} conducted extensive trials on numerous IBM devices to characterize crosstalk using simultaneous randomized benchmarking (SRB) \cite{47}. They concluded that not all couplings are sensitive to crosstalk, i.e., crosstalk between certain couplings is minimal, whereas crosstalk between some couplings can result in a 2X-3X increase in error rates, and crosstalk disappears after a 1-hop distance. The authors presented a gate scheduling strategy to minimize crosstalk based on crosstalk characterization results, where they serialized the parallel gates at the cost of greater program depth (run-time) and therefore decoherence. To examine the reduction in crosstalk error and rise in decoherence error due to gate serialization, the authors created an SMT-based scheduler (XtalkSched). Crosstalk-aware scheduling has been shown to enhance program integrity by up to 5.6X. Another paper in \cite{48} attempted dynamic qubit frequency assignment to reduce crosstalk. Qubits are frequency addressable, which means that if two adjacent qubits have sufficiently distinct operating frequencies, crosstalk from one will be reduced. However, the operating frequency range of a qubit is limited, resulting in frequency congestion. The authors suggested a software method that dynamically allocates separate frequencies to surrounding qubits to overcome the frequency crowding problem. When compared to the gate serialization technique, they find a 13.3X increase in program success rate. The solution is applicable to frequency tunable qubits, according to the authors.

\subsubsection{Leveraging extended native gates}
There are still very few gates that can be realized on quantum hardware. Only a single 2-qubit gate, the CNOT gate is supported in IBM’s superconducting harware and Mølmer–Sørensen gate in the IonQ’s trapped-ion hardware. However, new hardware is emerging which supports multiple 2-qubit gates \cite{49}\cite{50}\cite{51}. In general, reducing gate count is desired to minimize decoherence and gate error for better resilience. An extended native gate set can make the gate decomposition step more efficient. In \cite{49}, the authors note that a SWAP can be decomposed using two gates (1 CZ + 1 iSWAP) when both CZ and iSWAP gates are available as native instructions. However, if only CZ or iSWAP is available as the native gate, it takes 3 CZ/I SWAP to decompose a SWAP. As a result, in NISQ architectures with limited connectivity, an enlarged gate set can dramatically lower the gate count from SWAP insertions. For a test case on QAOA circuits, they report a 30$\%$ percent reduction in gate depth.

\subsubsection{Application Specific Compilation}
Quantum compilers typically use generic rules to optimize every given quantum program, and they don't take program-specific information into account while doing aggressive optimization. There have been recent papers \cite{52}\cite{53}\cite{54}, that give algorithm-specific compilation approaches for QAOA, which is an outstanding nearterm algorithm. The ZZ-interactions in QAOA (may be accomplished with 2 CNOTs and 1 RZ operation inside a level and are commutative, \cite{53} i.e. these operations can be reordered without affecting the circuit's output state. In \cite{53}, the authors propose several QAOA-specific optimizations, including parallelization of ZZ-operations using a binary bin-packing algorithm (Instruction Parallelization - IP), repeated compilation of QAOA-circuits with reordered layers guided by a branch-and-bound optimization heuristic (Iterative Compilation), and layer-by-layer circuit construction and compilation prioritizing operations that require fewer SWAPs (Incremental Compilation). They also discuss various techniques designed to alter QAOA-circuit features in order to execute intelligent initial qubit allocation (Qubit Allocation and Initial Mapping - QAIM/ Variation-aware Qubit Placement - VQP). Over the existing state-of-the-art methodologies, these techniques delivered a 53$\%$  reduction in circuit depth, a 23$\%$ reduction in gate count, and a 63$\%$ increase in estimated success probability of QAOA-circuits. In addition, the paper also demonstrated about 26$\%$ improvement in performance on an actual IBM device.

\section{Conclusion}
\label{sec:conclusion}
In this chapter, we introduced the basics of quantum computing architectures. We then covered various quantum hardware technologies which are at the center of research. We reviewed the quantum software stack with emphasis on compilation, mapping and optimization heuristics used for two of the most common quantum hardware – Superconducting Quantum computers and Trapped Ion quantum computers. In the last several years quantum computing has been experiencing a sustained period of growth. However, to make the promise of quantum computing a reality, we need innovations at different levels of the computing stack. Only improving hardware and inventing new quantum algorithms will not be sufficient, and we need a middle stack that will bridge the gap between algorithm and hardware. As full-fledged error correction will not be possible on NISQ devices, error mitigation is the way to move forward in the foreseeable future. Therefore, we went over the various types errors plaguing the quantum systems, their effect on program performance and reviewed recent literature pertaining to optimizing and mitigating the effect of noise on quantum program reliability.

\section{Acknowledgement}
\label{sec:ack}
This material is based upon work supported by NSF (CNS-1814710, DGE-1821766, CNS-2129675, CCF-2210963, DGE-2113839, ITE-2040667), gifts from Intel, seed grants from Penn State ICDS and Huck Institute of the Life Sciences.

\bibliographystyle{unsrt}
\bibliography{ref}

\end{document}